\documentclass[conference]{IEEEtran}
\IEEEoverridecommandlockouts
\usepackage{cite}
\usepackage{amsmath,amssymb,amsfonts}
\usepackage{algorithmic}
\usepackage[ruled,linesnumbered]{algorithm2e}
\usepackage{graphicx}
\usepackage{textcomp}
\usepackage{xcolor}
\usepackage[colorlinks]{hyperref}
\hypersetup{citecolor=blue}
\usepackage{balance}
\def\BibTeX{{\rm B\kern-.05em{\sc i\kern-.025em b}\kern-.08em
    T\kern-.1667em\lower.7ex\hbox{E}\kern-.125emX}}
\begin{document}
\title{Sparsity-Exploiting Channel Estimation For Unsourced Random Access With Fluid Antenna}

%\author{\IEEEauthorblockN{Zhentian Zhang\textsuperscript}
%	\IEEEauthorblockA{\textsuperscript{1}National Mobile Communications Research Laboratory,\\Frontiers Science Center for Mobile Information Communication and Security\\Southeast University, Nanjing 210096, P. R. China\\
%		\textsuperscript{2}Purple Mountain Laboratories, Nanjing 211111, P. R. China\\
%		Emails: \{zhangzhentian\}@seu.edu.cn}}

\author{\IEEEauthorblockN{Keru Zhou\textsuperscript{1}, Zhentian Zhang\textsuperscript{1}, Jian Dang\textsuperscript{1,2}, Qianqian Sun\textsuperscript{1}, Zaichen Zhang\textsuperscript{1,2}}
	\IEEEauthorblockA{\textsuperscript{1}National Mobile Communications Research Laboratory,\\Frontiers Science Center for Mobile Information Communication and Security\\Southeast University, Nanjing 210096, P. R. China\\
		\textsuperscript{2}Purple Mountain Laboratories, Nanjing 211111, P. R. China\\
		Emails: \{haibara, zhangzhentian, dangjian, sunqianqian, zczhang\}@seu.edu.cn}}
\maketitle

\begin{abstract}
This work explores the channel estimation (CE) problem in uplink transmission for unsourced random access (URA) with a fluid antenna receiver. The additional spatial diversity in a fluid antenna system (FAS) addresses the needs of URA design in multiple-input and multiple-output (MIMO) systems. We present two CE strategies based on the activation of different FAS ports, namely alternate ports and partial ports CE. Both strategies facilitate the estimation of channel coefficients and angles of arrival (AoAs). Additionally, we discuss how to refine channel estimation by leveraging the sparsity of finite scatterers. Specifically, the proposed partial ports CE strategy is implemented using a regularized estimator, and we optimize the estimator's parameter to achieve the desired AoA precision and refinement. Extensive numerical results demonstrate the feasibility of the proposed strategies, and a comparison with a conventional receiver using half-wavelength antennas highlights the promising future of integrating URA and FAS.
\end{abstract}

\section{Introduction}
Unsourced random access (URA) techniques \cite{WYP_Survey} address multiple access challenges in the context of massive machine-type communications with finite blocklengths. These techniques propose a potential multiple access code design to provide robust communication support. Significant degrees of freedom have been unlocked through its evolution towards a multiple-input and multiple-output (MIMO) system with multiple receiving antennas \cite{URA_China_Comm}. Recently, a novel MIMO system, the fluid antenna system (FAS), has garnered considerable attention due to its compelling features \cite{FAS}, \cite{Psomas-dec2023}, \cite{Khammassi-2023}. FAS offers additional spatial diversity with $N$ position reconfigurable antennas or ports spaced linearly \cite{New-twc2023}. In comparison to conventional $M$-antenna maximum ratio combining systems, FAS achieves better outage probability performance with a sufficient number of ports. As a result, FAS channel reconstruction becomes essential for selecting the ports with the best channel gains \cite{FAS_Turorial}.

Therefore, URA will naturally leverage the diversity gain generated by FAS, as the precision of channel estimation (CE) is pivotal for obtaining desirable decoding results at the MIMO URA receiver. Specifically, while URA often emphasizes promising concatenated encoder/decoder designs, the CE performance directly affects the performance of concatenated decoding under MIMO channels. Consequently, we investigate potential CE strategies for FAS-URA in this paper. The main focuses of this paper are as follows:
\subsubsection{Channel Estimation For FAS-URA} 
Under limited blocklengths and finite radio-frequency (RF) chains, we explain how to perform activity detection (AD) and estimate channel coefficients, considering the FAS planar geometric model in \cite{FAS_geometric channel}, which includes both scattered and line-of-sight (LOS) paths. We propose two strategies, namely alternate ports and partial ports CE, to facilitate these estimations. Additionally, we enable the estimation of arrival angles (AoAs) using a channel dictionary, exploiting the sparsity of finite scatterers.
\subsubsection{Partial Ports Estimation With Regularized Estimator}
We propose a regularized estimator for partial ports estimation, with or without prior knowledge of activity. Building on this regularized estimator, we further formulate the problem of port selection as an NP-hard problem with a feasible solution, providing guidance on how to achieve better AoA estimation.
\subsubsection{Comparison With Conventional Uniform Linear Array (ULA)}
Specifically, we compare the estimation performance between the proposed FAS-URA and the conventional ULA with the half-wavelength rule \cite{Quantized_problem, ULA_geometric}, which has been widely considered in current URA designs \cite{ULA_URA1, ULA_URA2}. Numerical results demonstrate that the proposed CE via alternate ports provides better AoA precision, while the proposed CE via partial ports facilitates more accurate channel construction compared to the conventional ULA.

In the following, the system model for uplink FAS-URA is presented in Sec.\ref{sec.2}. In Sec.\ref{sec.3}, the proposed alternate ports CE is discussed. Sec.\ref{sec.4} introduces the proposed AoA estimation using a channel dictionary, leveraging the sparsity of finite scatterers. In Sec.\ref{sec.5}, the proposed partial ports CE is detailed, along with the regularized estimator and the ports index selection optimization problem. Numerical results are provided in Sec.\ref{sec.6}, and the conclusion is presented in Sec.\ref{sec.7}.
\section{URA With Fluid Antenna System Model}\label{sec.2}
\subsection{Uplink Transmission Setups}
We assume an uplink transmission with $T_p$ duration where there are $K_a$ single traditional antenna devices to be served by a receiver with a $W\lambda$-wavelength fluid antenna. The fluid antenna has $N$ positions/ports to switch\footnote{The potential switching delay becomes negligible with advances in antenna designs and novel materials\cite{FAS},\cite{FAS4},\cite{FAS2}.} and is connected to $M$ RF-chains where normally $M \ll N$, i.e., only signals received from $M$ ports connected to RF-chains can be de-mixed from analogue domain. Quasi-static fading without asynchronous error is considered in this work\footnote{All works in this paper is readily applicable under scenario with large fading effect considering path loss.}, i.e., the channel coefficients between devices and receiver retains unchanged during the transmission. For transmission in the URA\cite{WYP_Survey,URA_China_Comm}, the $k$-th device, $k\in [1:K_a]$ uses $B_p$ bits to select an $L_p$-length ($L_p\le T_p$) pilot/codeword $\boldsymbol{x}_k \in \mathbb{C}^{L_p \times 1}$ from the common codebook $\boldsymbol{A} \in \mathbb{C}^{L_p \times 2^{B_p}}$ which can be generated from Gaussian signatures or sub-sampled DFT matrx\cite{Collision_Probability}. And the pilot follows the power constraint of energy-per-bit under finite duration by $E_b/N_0=\left(E_sT_p\right)/\left(B_p\sigma^2\right)$ where $\|\boldsymbol{x}_k\|_2^2 = E_sL_p$ and $E_s$ is the unit power per channel uses.
\subsection{Planar Propagation URA Channel Models: FAS vs ULA}
For ease of description and the integrity of the content, we elaborate the geometric models of FAS and ULA under planar propagation\footnote{Near-filed problem is not considered in this work.}. FAS generates diversity by allowing the existence of multiple receiving points within fixed antenna length while the traditional ULA has a \textit{half-wavelength} rule to follow. Thereupon, the following assumes both antennas have identical physical antenna size, i.e., $W\lambda=(M-1)\frac{\lambda}{2}$ with half-wavelength rule. Also, we consider transmission with finite $L_s$ scatterers.
%\footnote{The assumption that the number of scatterers is a known prior can be justified if some mature techniques are used in advance, such as the multiple signal classification (MUSIC) algorithm for azimuth AoAs.}

\begin{figure}[htp]
	\centering
	\includegraphics[width=3.4in]{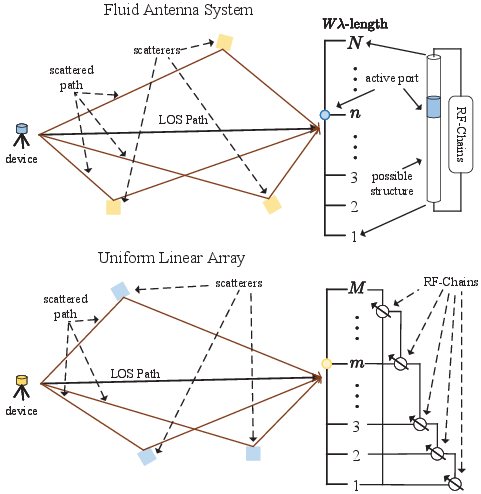}
	\caption{Illustration of planar propagation with LOS path and multiple scattered paths under possible FAS and ULA structures. Only finite RF-chains are assumed, i.e., $M\ll N$.}
	\label{geometric}
\end{figure}

In Fig. \ref{geometric}, the planar propagation geometric model is depicted with finite scatterer paths and LOS path. The pivotal feature of FAS compared with traditional ULA is that the gap between $N$ ports no longer follows the half-wavelength rule\footnote{This means there will be many correlated ports in vicinity, which however introduces more degree-of-freedom. Please refer to \cite{FAS3},\cite{FAS_Turorial} for more detail.}, i.e., for ULA, the gap distance $d$ between adjacent units equals to $d=\frac{\lambda}{2}$ while the gap distance is $d=\frac{W\lambda}{N-1}$ for FAS. Let $\boldsymbol{g}_k$ denote the channel coefficient vector of the $k$-th device. However, both antennas always abide by identical physical size and share identical number of functional RF-chains.

For FAS\footnote{Fig. \ref{geometric} is an abstract illustration of the FA structure which does not imply any specific implementation method. It could be realized by metamaterial or reconfigurable pixel designs if fast switching speed is required. Also, a two port configuration for FA is also possible.} channel\cite{FAS_geometric channel} $\boldsymbol{g}_k=\left[g_{k,1},g_{k,2},\ldots,g_{k,N}\right]^{\mathrm{T}} \in\mathbb{C}^{N\times 1}$, element $g_{k,n}, n\in[1:N]$ with the first port as benchmark can be expressed as:
\begin{equation}
	\label{eq:1}
	g_{k,n}=\underbrace{\sigma_{k,0}e^{-j\frac{2\pi(n-1)W}{N-1}\cos\theta_{k,0}}}_{\text{LOS component}}+\underbrace{\sum_{l=1}^{L_s}\sigma_{k,l}e^{-j\frac{2\pi(n-1)W}{N-1}\cos\theta_{k,l}}}_{\text{scatterers component}},
\end{equation}
where  $\Omega$ is the average energy of the channel, $L_s$ is the number of scattered paths. For LOS component, $\sigma_{k,0}=\sqrt{\frac{K\Omega}{K+1}}e^{j\alpha_k}$, $K$ is the Rice factor, $\alpha$ is the arbitrary LOS phase, and $\theta_{k,0}$ represents the azimuth AoA of the specular component of the $k$-th device. For scatterers component, $\sigma_{k,l}$ is the complex coefficient of the $l$-th path of the $k$-th device, $\theta_{k,l},l\in[1:L_s]$ represents the azimuth AoAs of the $l$-th path of $k$-th device, and $\sigma_{k,l}$ is the complex coefficient of the $l$-th path of $k$-th device satisfying $\sum_{l=1}^{L_s}|\sigma_{k,l}|^2=\frac{\Omega}{K+1}$. Moreover, for the traditional ULA geometric model\cite{ULA_geometric,Quantized_problem}, $\boldsymbol{g}_k=[g_{k,1},g_{k,2},\ldots,g_{k,M}]^{\mathrm{T}}\in\mathbb{C}^{M\times 1}$, element $g_{k,m},m\in[1:M]$ can be expressed as
\begin{equation}
	\label{eq:2}
	g_{k,m}=\underbrace{\sigma_{k,0}e^{-j\frac{2\pi(m-1)d}{\lambda}\cos\theta_{k,0}}}_{\text{LOS component}}+\underbrace{\sum_{l=1}^{L_s}\sigma_{k,l}e^{-j\frac{2\pi(m-1)d}{\lambda}\cos\theta_{k,l}}}_{\text{scatterers component}},
\end{equation}
where only the spatial response part $e^{-j\frac{2\pi(m-1)d}{\lambda}}$ is different from FAS and others share identical physical representations in \eqref{eq:1}. Yet, we have to note that there is no way the FAS receiver can observe decoupled signals from all $N$ ports simultaneously due to the limited RF-chains $M \ll N$. And we have made fair assumptions including the length of antenna and the number of RF-chains. In the sequel, the work is established on the justified assumptions with adequate equity.
\begin{figure*}[htp]
	\centering
	\includegraphics[width=4.8in]{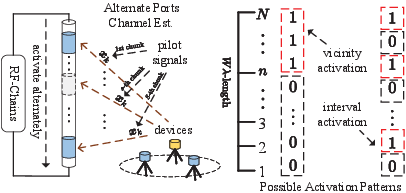}
	\caption{Illustration of the proposed AP-CE strategy and possible ports activation patterns (vicinity/interval activation).}
	\label{AP-CE}
\end{figure*}
\section{Alternate Ports Channel Estimation}\label{sec.3}
Reminding that only $M$ ports can be activated simultaneously, however, the receiver can make proper ports selection only with knowledge of all channel coefficients between ports and devices. Based on this observation, an alternate ports channel estimation (AP-CE) strategy is proposed for FA-URA. Specifically, AP-CE receives pilot signals from different sets of ports in each duration of time while switching different sets of ports to be activated. In this way, the receiver can observe the channel condition of all ports. In Fig. \ref{AP-CE}, we illustrate the proposed AP-CE strategy and give examples on possible ports activation patterns including activating sets of ports in vicinity or ports in interval gaps every switch time.

Let $N_{obs},N_{obs}\le M$ denote the number of ports activated/observed by the receiver. Therefore, there are $S=N/N_{obs}$ sets of ports to be estimated and the duration for CSI estimation is also split into $S$ chunks with channel uses $T_p=SL_p$. The channel vector between the $k$-th device and all ports based on model in \eqref{eq:1} can be written as:
\begin{equation}
	\label{eq:3}
	\boldsymbol{g}_k=
	\begin{bmatrix}
		g_{k,1}\\
		g_{k,2}\\
		g_{k,3} \\
		\vdots\\
		g_{k,n}\\
		\vdots\\
		g_{k,N}\\
	\end{bmatrix}
	=\begin{bmatrix}
		\boldsymbol{g}_{k,1}\\
		\boldsymbol{g}_{k,2}\\
		\vdots\\
		\boldsymbol{g}_{k,s}\\
		\vdots\\
		\boldsymbol{g}_{k,S}\\
	\end{bmatrix},
	\underbrace{\boldsymbol{g}_{k,s}}_{s\in[1:S]}=\begin{bmatrix}
				g_{k,(1+(s-1)N_{obs})}\\
		g_{k,(2+(s-1)N_{obs})}\\
		g_{k,(3+(s-1)N_{obs})}\\
		\vdots \\
		g_{k,(sN_{obs})}\\
	\end{bmatrix},
\end{equation}
where $\boldsymbol{g}_{k,s}\in \mathbb{C}^{N_{obs}\times 1}$ is the channel coefficient between the activated ports and $k$-th device at the $s$-th chunk of duration time where there are $N_{obs}$ elements to be estimated. During each chunk of duration, the received signal $\boldsymbol{Y}_{p,s}\in\mathbb{C}^{L_p \times N_{obs}}$ can be written as:
\begin{equation}
	\label{eq:4}
	\boldsymbol{Y}_{p,s}= \sum_{k}\boldsymbol{x}_k\boldsymbol{g}_{k,s}^{\mathrm{T}} + \boldsymbol{N}_{s},s\in[1:S],
\end{equation}
where $\boldsymbol{N}_{s}$ is the additive white Gaussian noise (AWGN) following distribution of $\mathcal{CN}(\boldsymbol{0},\sigma^2\boldsymbol{I})$ with zero mean and $\sigma^2$ variance. Therefore, the total signals $\boldsymbol{Y}_{p}\in \mathbb{C}^{L_p\times N}$ received in $T_p$ duration is
\begin{equation}
	\label{eq:5}
	\begin{aligned}
		\boldsymbol{Y}_{p}&=\underbrace{[\boldsymbol{Y}_{p,1},\boldsymbol{Y}_{p,2},\ldots,\boldsymbol{Y}_{p,S}]}_{\text{$S$ chunks}} \\
		&=\sum_{k}\boldsymbol{x}_k^p\boldsymbol{g}_k^{\mathrm{T}}+\boldsymbol{N}=\boldsymbol{A}\boldsymbol{G}+\boldsymbol{N},
	\end{aligned}
\end{equation}
where $\boldsymbol{N}$ represents AWGN, $\boldsymbol{G}\in \mathbb{C}^{2^{B_p}\times N}$ is a row-sparse matrix with only nonzero elements in $K_a$ rows and others all zeros and $\boldsymbol{A}$ is the common codebook. Thus, the joint AD and CE problem in \eqref{eq:5} is formulated into a sparse recovery problem. 

However, we have to note that model in \eqref{eq:5} is distinctive to the common linear model under conventional ULA channel which is normally in the form of $\boldsymbol{Y}=\boldsymbol{A}\boldsymbol{H}+\boldsymbol{N}$ where $\boldsymbol{H}\in\mathbb{C}^{2^{B_p}\times M}$. Since ULA has $M$ antenna unit all connected to RF-chains, the receiver with ULA antenna can receive decoupled signals from all $M$ unit throughout the $T_p$ duration, i.e., $L_p=T_p$ while for the proposed AP-CE, $L_p=T_p/N$ and $M\ll N$. Though the pilot length is shortened in FA-URA, the diversity of antenna is increased under FA-URA. We solve this sparse recovery problem following the spirit of simultaneous orthogonal matching pursuit (SOMP)\cite{SOMP}. We summarize the SOMP in Algorithm \ref{alg:SOMP_MIMO}.
\begin{algorithm}[htp]
	\SetAlgoLined
	\label{alg:SOMP_MIMO}
	\KwIn{Signal $\boldsymbol{Y}_p$, common codebook $\boldsymbol{A}$}
	\KwOut{Active Indices and Estimated Channel Vectors}
	Initialization: $k\leftarrow 0$, residual $\boldsymbol{R}\leftarrow \boldsymbol{Y}_p$, $\mathcal{S}\leftarrow \emptyset$, $\boldsymbol{G}\leftarrow \boldsymbol{0}$ \;
	\While{$k \le \tilde{K}_a$}{
		$i\leftarrow \mathop{\arg\max}\limits_{i_k^p\in\{1,2,\ldots,2^{B_p}\}}\|\boldsymbol{R}^{\mathrm{H}}\boldsymbol{A}(:,i_k^p)\|_2/\|\boldsymbol{A}(:,i_k^p)\|_2$\;
		Activity Support $\mathcal{S}\leftarrow \mathcal{S} \cup i$\;
		Projection Span Space $\boldsymbol{\Phi}\leftarrow \boldsymbol{A}\left(:,\{\mathcal{S}\}\right)$ \;
		Update residual $\boldsymbol{R}\leftarrow \left(\boldsymbol{R}-\boldsymbol{\Phi}\boldsymbol{\Phi}^{\dagger}\boldsymbol{Y}_p\right )$ and $\left(\cdot\right)^{\dagger}$ denotes Moore-Penrose inverse\;
		$k\leftarrow k+1$\;}
	Active support $\mathcal{S}$ and 
	channel estimation $\{\tilde{\boldsymbol{g}}_k\}\leftarrow \left(\boldsymbol{A}\left(:,\{\mathcal{S}\}\right)^{\mathrm{H}}\boldsymbol{A}\left(:,\{\mathcal{S}\}\right)\right)^{-1}\boldsymbol{A}\left(:,\{\mathcal{S}\}\right)^{\mathrm{H}}\boldsymbol{Y}_p.$
	\caption{AP-CE By SOMP}
\end{algorithm}
\begin{figure*}[htp]
	\centering
	\includegraphics[width=4.5in]{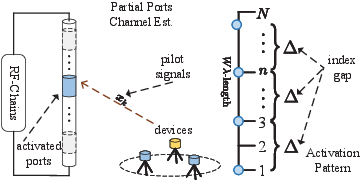}
	\caption{Illustration of the proposed PP-CE strategy and ports activation patterns with index gap $\Delta$.}
	\label{regularized}
\end{figure*}
SOMP is a greedy iterative algorithm to solve sparse recovery problems. SOMP algorithm selects the column of the measurement matrix with the maximum correlation between the current residual at each iteration and adds its index to the output list. The complexity of SOMP is dominated by the Moore-Penrose inverse at each iteration, scaling as $\mathcal{O}(NL_p^3)$. Meanwhile, the line 2 of SOMP algorithm requires the number of activity which can be estimated by the power information of the received signal:
\begin{equation}
	\label{eq:6}
	\tilde{K}_{a} =  \left \lfloor \frac{1}{E_sL_p}\left( \frac{ \left \| \boldsymbol{Y}_p \right \|_{F}^2 }{N\Omega}-\sigma^2L_p \right) \right \rceil.
\end{equation}
\section{AoA Estimation And Channel Refinement}\label{sec.4}
In accordance with model \eqref{eq:1} and \eqref{eq:3}, it is possible to estimate AoAs and even the complex coefficients of the propagation paths, which is often known as channel reconstruction. Let $I_{a}$ denote the number of AoA samples and is adequate large for certain AoA resolution. One can form a channel dictionary based on \eqref{eq:1}. The normalized ports response vector $\boldsymbol{d}_i \in \mathbb{C}^{N\times 1}$ at the receiver can be written as:
\begin{equation}
	\label{eq:7}
	\boldsymbol{d}_i=\frac{1}{\sqrt{N_{obs}}}\left[1,e^{-j\frac{2\pi W}{N-1}\cos\theta_i},\ldots,e^{-j\frac{2\pi(n-1)W}{N-1}\cos\theta_i},\ldots\right]^{\mathrm{T}}
\end{equation}
where $\theta_i, i\in [1:I_a]$ is the AoA sample, for alternate ports estimation $N_{obs}=N$, for partial ports estimation, $N_{obs}=M$. Therefore, one can obtain a prescribed channel response dictionary $\boldsymbol{D}=\left[\boldsymbol{d}_1,\boldsymbol{d}_2,\ldots,\boldsymbol{d}_{I_a}\right]\in \mathbb{C}^{N \times I_a}$. Meanwhile, considering there are $(L_s+1)$ propagation paths (if LOS is considered), we can write \eqref{eq:1} in a compact form:
\begin{equation}
	\label{eq:8}
	\begin{aligned}
		\boldsymbol{g}_k &= \sum_{l=0}^{L_s}\sigma_{k,l}\boldsymbol{d}_{l} \\
		&=\underbrace{\begin{bmatrix}
				\boldsymbol{d}_0 & \ldots & \boldsymbol{d}_l & \ldots &\boldsymbol{d}_{L_s}
		\end{bmatrix}}_{\text{ports response}}
		\underbrace{\begin{bmatrix}
				\sigma_{k,0}\\
				\vdots\\
				\sigma_{k,l}\\
				\vdots\\
				\sigma_{k,L_s}
		\end{bmatrix}}_{\text{path coefficients}}
	\end{aligned} =\boldsymbol{D}\boldsymbol{\sigma}_{k},
\end{equation}
where $\boldsymbol{\sigma}_{k} \in \mathbb{C}^{I_a \times 1}$ is an $(L_s+1)$-sparse vector denoting the path coefficient vector. Thus, AoA estimation is also converted into a sparse recovery problem. We note that the SOMP is also applicable for model \eqref{eq:8}. The estimated channel vectors by Algorithm \ref{alg:SOMP_MIMO} can be refined by the AoA dictionary.
%\subsection{Ports Selection Criteria}
%After channel estimation refinement, it common to determine which ports are with the best quality which is the feature compared with traditional systems. To achieve this goal, we first calculate the signal to interference-plus-noise ratio (SINR) at each ports of devices and use the overall sum-rate approximation as ports selection criteria.
%\begin{equation}
%	\label{eq:9.1}
%	\begin{aligned}
%		\gamma_{k,n} &= \frac{|\tilde{g}_{k,n}|^2\cdot\|\boldsymbol{x}_k\|_2^2}{\sigma^2L_p+\sum\limits_{i\neq k} |\tilde{g}_{k,n}|^2\cdot\|\boldsymbol{x}_k\|_2^2}\\
%		&=\frac{|\tilde{g}_{k,n}|^2}{\sigma^2/E_s+\sum\limits_{i\neq k}|\tilde{g}_{k,n}|^2},
%	\end{aligned}
%\end{equation} 
%where $\tilde{g}_{k,n}$ is the $n$-th estimated channel coefficient of the $k$-th codeword. The overall sum-rate can then be calculated as:
%\begin{equation}
%	\label{eq:9.2}
%	C = \sum_{k=1}^{K_a}\log\left(1+\sum_{n=1}^{N_{obse}}\gamma_{k,n}\right).
%\end{equation}

\section{Partial Ports Channel Estimation}\label{sec.5}
Another plausible way is only observing signals from $N_{obs}\le M$ ports within the $T_p$ duration, i.e., the length of pilot is $L_p=T_p$ and no longer shortened which is called as partial ports channel estimation (PP-CE). Therefore, the received signal model resembles \eqref{eq:5} except for that $Y_p\in\mathbb{C}^{T_p \times N_{obs}}$, $\boldsymbol{A}\in\mathbb{C}^{T_p \times 2^{B_p}}$, $\boldsymbol{G}\in\mathbb{C}^{2^{B_p} \times N_{obs}}$. We illustrate the PP-CE in Fig. \ref{regularized}. The pilot transmission is no longer split into chunks. The aforementioned methods in Sec. III and IV can be still utilized to conduct AD, CE and AoA estimation. However, in order to reconstruct the channel from partially observed ports. The receiver needs to select good ports to be observed. The following elaborates how to choose good $\Delta$.

Assuming ports are observed from the 1st port with index gap $\Delta$. Based on \eqref{eq:7},\eqref{eq:8}, the coefficient of the observed ports can be expressed as:
\begin{equation}
	\label{eq:9}
	\begin{aligned}
	&\boldsymbol{g}_k=\begin{bmatrix}
		g_{k,1}\\
		g_{k,2}\\
		\vdots\\
		g_{k,N_{obs}}
	\end{bmatrix}
	=
	\begin{bmatrix}
		\sum_{l=0}^{L_s}\sigma_{k,l} \\
		\sum_{l=0}^{L_s}\sigma_{k,l} e^{-j\frac{2\pi \Delta W}{N-1}\cos\theta_{k,l}}\\
		\vdots\\
		\sum_{l=0}^{L_s}\sigma_{k,l} e^{-j\frac{2\pi N_{obs}\Delta W}{N-1}\cos\theta_{k,l}}
	\end{bmatrix}= \\
	&\underbrace{\begin{bmatrix}
		1 &   \ldots & 1\\
		e^{-j\frac{2\pi \Delta W}{N-1}\cos\theta_{k,0}} & \ldots & e^{-j\frac{2\pi \Delta W}{N-1}\cos\theta_{k,L_s}}\\
		\vdots &  \ddots  & \vdots\\
		e^{-j\frac{2\pi \Delta N_{obs}W}{N-1}\cos\theta_{k,0}}  & \ldots & e^{-j\frac{2\pi \Delta N_{obs}W}{N-1}\cos\theta_{k,L_s}}
	\end{bmatrix}
	\begin{bmatrix}
		\sigma_{k,0}\\
		\sigma_{k,1}\\
		\vdots\\
		\sigma_{k,L_s}
	\end{bmatrix}}_{\boldsymbol{W}\boldsymbol{\sigma}_{k}},
	\end{aligned}
\end{equation}
where $\boldsymbol{W}\in \mathbb{C}^{N_{obs}\times (L_s+1)}$ is a Vandermonde matrix with $\Delta$ as parameter to be optimized. Moreover, it requires $N_{obs}\ge L_s+1$ to estimate the path coefficients. Without loss of generality, the relationship between path coefficients and the estimated channel is $\tilde{\boldsymbol{g}}_k= \boldsymbol{W}\boldsymbol{\sigma}_{k}+\boldsymbol{n}_k$ where $\boldsymbol{n}_k$ is noisy derivations with variance $\sigma^2_d$. The goal is to obtain the path coefficients with minimum deviation, we form the subject goal as:
\begin{equation}
	\label{eq:10}
	\mathop{\arg\min}\limits_{\boldsymbol{\sigma}_k} \|\tilde{\boldsymbol{g}}_k-\boldsymbol{W}\boldsymbol{\sigma}_k\|_2^2 +\underbrace{\frac{(K+1)\alpha}{\Omega}\|\boldsymbol{\sigma}_k\|_2^2}_{\text{regularizer with $\gamma=\frac{(K+1)\alpha}{\Omega}$}},
\end{equation}
where the regularizer is considered due to aggregate path coefficient power in \eqref{eq:1} and regularization constant is flexibly parameterized. The target function can be formulated as:
\begin{equation}
	\label{eq:11}
	\begin{aligned}
		&\|\tilde{\boldsymbol{g}}_k-\boldsymbol{W}\boldsymbol{\sigma}_k\|_2^2 +\gamma\|\boldsymbol{\sigma}_k\|_2^2\\
		=&(\tilde{\boldsymbol{g}}_k-\boldsymbol{W}\boldsymbol{\sigma}_k)^{\mathrm{H}}(\tilde{\boldsymbol{g}}_k-\boldsymbol{W}\boldsymbol{\sigma}_k)+\gamma\boldsymbol{\sigma}_k^{\mathrm{H}}\boldsymbol{\sigma}_k\\
		=&\tilde{\boldsymbol{g}}^{\mathrm{H}}\tilde{\boldsymbol{g}}-\tilde{\boldsymbol{g}}^{\mathrm{H}}\boldsymbol{W}\boldsymbol{\sigma}_k-\boldsymbol{\sigma}_k^{\mathrm{H}}\boldsymbol{W}^{\mathrm{H}}\tilde{\boldsymbol{g}}+\boldsymbol{\sigma}_k^{\mathrm{H}}\boldsymbol{W}^{\mathrm{H}}\boldsymbol{W}\boldsymbol{\sigma}_k+\gamma\boldsymbol{\sigma}_k^{\mathrm{H}}\boldsymbol{\sigma}_k,
	\end{aligned}
\end{equation}
whose gradient by $\boldsymbol{\sigma}_k$ can be calculated as
\begin{equation}
	\label{eq:12}
	\nabla_{\boldsymbol{\sigma}^2}=-2\boldsymbol{W}^{\mathrm{H}}\tilde{\boldsymbol{g}}_k+2\boldsymbol{W}^{\mathrm{H}}\boldsymbol{W}\boldsymbol{\sigma}_k+2\gamma\boldsymbol{\sigma}_k,
\end{equation}
and by setting $\nabla_{\boldsymbol{\sigma}^2}=0$, the solution of \eqref{eq:10} is:
\begin{equation}
	\label{eq:13}
	\tilde{\boldsymbol{\sigma}}_k=(\boldsymbol{W}^{\mathrm{H}}\boldsymbol{W}+\gamma\boldsymbol{I})^{-1}\boldsymbol{W}^{\mathrm{H}}\tilde{\boldsymbol{g}}_k,
\end{equation}
where $\boldsymbol{I}$ is the unit matrix. With activity results, one can estimate channels by \eqref{eq:13} with regularized estimator. Subsequently, the goal is to find out $\Delta$ producing lowest estimation error, the deviation of path coefficients can be expressed as:
\begin{equation}
	\label{eq:14}
	\begin{array}{cl}
		\mathop{\arg\min}\limits_{\Delta}\|\tilde{\boldsymbol{\sigma}}_k-\boldsymbol{\sigma}_k\|_2^2 \\
		\text { subject to } \Delta\in \mathcal{N}_0=\{1,2,\ldots,\frac{N}{M}\},
	\end{array}
\end{equation}
Substitute \eqref{eq:9} into \eqref{eq:14}, we have:
\begin{equation}
	\label{eq:15}
	\begin{aligned}
		&\|\tilde{\boldsymbol{\sigma}}_k-\boldsymbol{\sigma}_k\|_2^2\\
		=&\|(\boldsymbol{W}^{\mathrm{H}}\boldsymbol{W}+\gamma\boldsymbol{I})^{-1}\boldsymbol{W}^{\mathrm{H}}(\tilde{\boldsymbol{g}}_k-\boldsymbol{g}_k) \|_2^2 \\
		=&\|(\boldsymbol{W}^{\mathrm{H}}\boldsymbol{W}+\gamma\boldsymbol{I})^{-1}\boldsymbol{W}^{\mathrm{H}} \boldsymbol{n}_k \|_2^2 \\
	\end{aligned}
\end{equation}
By singular value decomposition, we have decomposed matrix $\boldsymbol{W}=\boldsymbol{U}\boldsymbol{\Sigma}\boldsymbol{V}^{\mathrm{H}}$ where $\boldsymbol{U},\boldsymbol{V}$ are unitary matrices and unitary matrix has favourable features such as $\boldsymbol{U}^{\mathrm{H}}\boldsymbol{U}=\boldsymbol{I}$, $\boldsymbol{U}^{\mathrm{H}}=\boldsymbol{U}^{-1}$, and $\|\boldsymbol{U}\boldsymbol{x}\|_2^2=\|\boldsymbol{x}\|_2^2$. Substitute decomposed $\boldsymbol{W}$ into \eqref{eq:15}:
\begin{equation}
	\label{eq:16}
	\begin{aligned}
		&\|(\boldsymbol{W}^{\mathrm{H}}\boldsymbol{W}+\gamma\boldsymbol{I})^{-1}\boldsymbol{W}^{\mathrm{H}} \boldsymbol{n}_k \|_2^2 \\
		=&\| ( ( \boldsymbol{V}\boldsymbol{\Sigma}^{\mathrm{H}}\boldsymbol{U}^{\mathrm{H}} )(\boldsymbol{U}\boldsymbol{\Sigma}\boldsymbol{V}^{\mathrm{H}}) +\gamma\boldsymbol{I})^{-1} (\boldsymbol{V}\boldsymbol{\Sigma}^{\mathrm{H}}\boldsymbol{U}^{\mathrm{H}})\boldsymbol{n}_k  \|_2^2 \\
		=&\|  ( \boldsymbol{V}\boldsymbol{\Sigma}^{\mathrm{H}}\boldsymbol{\Sigma}\boldsymbol{V}^{\mathrm{H}}+\gamma\boldsymbol{V}\boldsymbol{V}^{\mathrm{H}} )^{-1}(\boldsymbol{V}\boldsymbol{\Sigma}^{\mathrm{H}}\boldsymbol{U}^{\mathrm{H}})\boldsymbol{n}_k\|_2^2 \\
		=&\| ( \boldsymbol{\Sigma}^{\mathrm{H}}\boldsymbol{\Sigma+\gamma\boldsymbol{I}})^{-1}\boldsymbol{\Sigma}^{\mathrm{H}}\boldsymbol{n}_k \|_2^2 \\
		\approx&\sigma_c^2\mathrm{Tr}\left[ ((\boldsymbol{\Sigma}^{\mathrm{H}}\boldsymbol{\Sigma}+\gamma\boldsymbol{I})^{-1}\boldsymbol{\Sigma}^{\mathrm{H}})^2 \right]
	\end{aligned}
\end{equation}
 where $\mathrm{Tr}[\cdot]$ denotes the trace of the matrix. Hereafter, the problem in \eqref{eq:14} is now equivalent to:
 \begin{equation}
 	\label{eq:17}
 	\begin{array}{cl}
 		\mathop{\arg\min}\limits_{\Delta}\mathrm{Tr}\left[( (\boldsymbol{\Sigma}^{\mathrm{H}}\boldsymbol{\Sigma}+\gamma\boldsymbol{I})^{-1}\boldsymbol{\Sigma}^{\mathrm{H}})^2 \right] \\
 		\text { subject to } \Delta\in \mathcal{N}_0=\{1,\ldots,\frac{N}{M}\},
 	\end{array}
 \end{equation}
which is an NP-hard problem. However, we can solve this problem via exhaust search since $|\mathcal{N}_0|=\frac{N}{M} \ll N$ with maximum complexity $\mathcal{O}(\frac{N}{M} )$.
\section{Numerical Results}\label{sec.6}
In this section, we illustrate the results of the proposed AP-CE and the proposed PP-CE with regularized estimator. We compare the results with URA CE under ULA model. Universal parameters are $K_a=5$, amount of finite scatters $L_s=3$, $T_p=200$, Rive factor $K=2$, $\Omega=1$, number of AoA samples $I_a=100$, number of RF-chains and ports $M=10$ and $N=100$. For AP-CE, $N_{obs}=N$ and for PP-CE, $N_{obs}=M$. The averaged normalized mean square error (NMSE) of CE is defined by $\mathbb{E}\{\| \boldsymbol{g}_k-\tilde{\boldsymbol{g}}_k \|_2^2 \} / \mathbb{E}\{\| \boldsymbol{g}_k\|_2^2\}$ and the precision of AoA estimation is denoted by AoA NMSE $\mathbb{E}\{| \theta_{k,l}-\tilde{\theta}_{k,l} | \} / \mathbb{E}\{| \theta_{k,l}|\}$. If $K_a$ is known, the AD error consists of only error of missed detection otherwise AD error is the summation of error of missed detection and false alarm. Only those correctly detected are used to calculate the corresponding NMSE.
\begin{figure}[htp]
	\centering
	\includegraphics[width=3.8in]{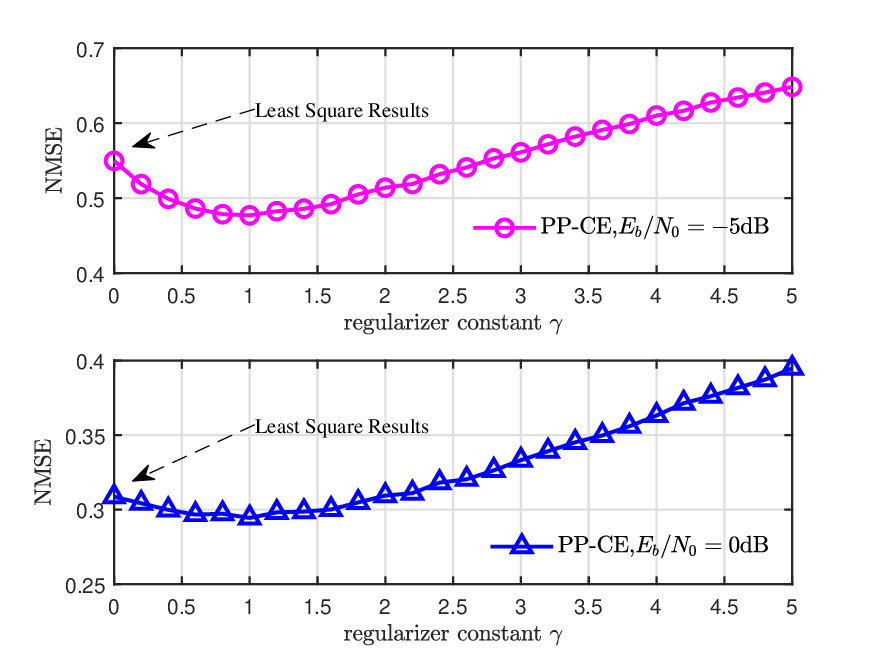}
	\caption{Performance with different power allocation under geometric model.}
	\label{regularizer_constant}
\end{figure} 
\subsubsection{PP-CE Parameter Setups $\gamma$ and $\Delta$}
We illustrate the performance of the proposed PP-CE with different regularizer $\gamma$ and the value of \eqref{eq:17} for justified parameter setups. 
Firstly, we illustrate the channel NMSE versus $\gamma$ with $\Delta=10$ in Fig. \ref{regularizer_constant}. The viability of the proposed regularized CE is validated compared with the results under $\gamma=0$ which the least square-oriented estimation referred in \cite{FAS_Turorial}. Improved CE can be observed and the following simulations are conducted with $\gamma=1$.
\begin{figure}[htp]
	\centering
	\includegraphics[width=3.5in]{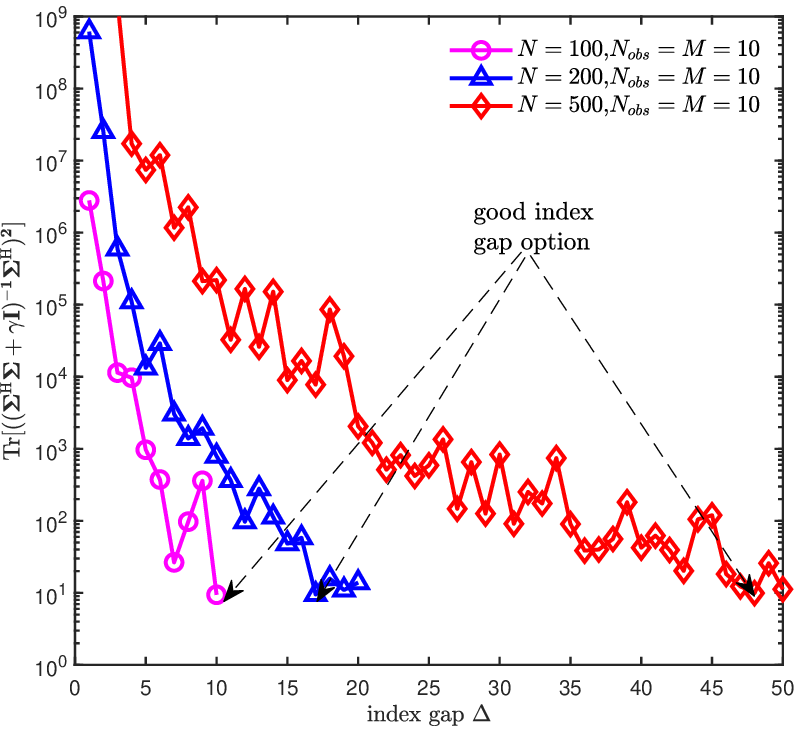}
	\caption{The value of subject function \eqref{eq:17} under different index gaps and number of ports.}
	\label{Index_gap_selection}
\end{figure}

\begin{figure}[htp]
	\centering
	\includegraphics[width=3.5in]{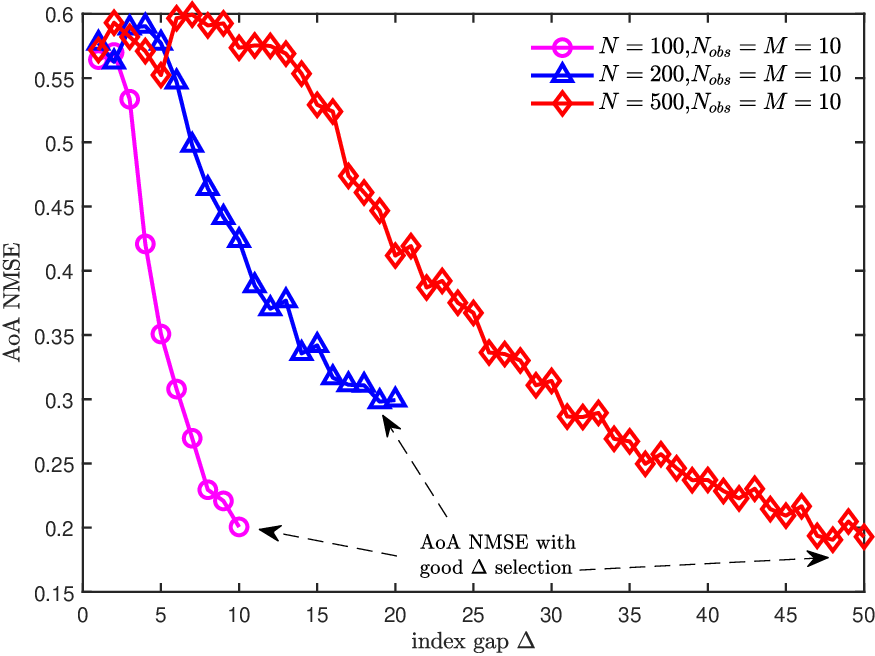}
	\caption{Performance of AoA estimation of PP-CE versus index gap under different $N$ and $E_b/N_0=0$dB.}
	\label{AoA_NMSE_with_index_gap}
\end{figure}

Secondly, we illustrate the value of target function $\mathrm{Tr}\left[( (\boldsymbol{\Sigma}^{\mathrm{H}}\boldsymbol{\Sigma}+\gamma\boldsymbol{I})^{-1}\boldsymbol{\Sigma}^{\mathrm{H}})^2 \right]$ under different $\Delta$ to determine good index gap option. Fig. \ref{Index_gap_selection} depicts the subjection function value versus different $\Delta$ and $N$ with $E_b/N_0=0$dB. In the sequel, index gap is fixed to $\Delta=10$. To further validate the effectiveness of the proposed gap selection strategy, we illustrate the AoA NMSE versus different $\Delta$ in Fig. \ref{AoA_NMSE_with_index_gap}. Similar tenancy can be observed in Fig. \ref{Index_gap_selection} and \ref{AoA_NMSE_with_index_gap} indicating the viability of the proposed selection strategy.

\subsubsection{AD and AoA Estimation}
\begin{figure}[htp]
	\centering
	\includegraphics[width=3.5in]{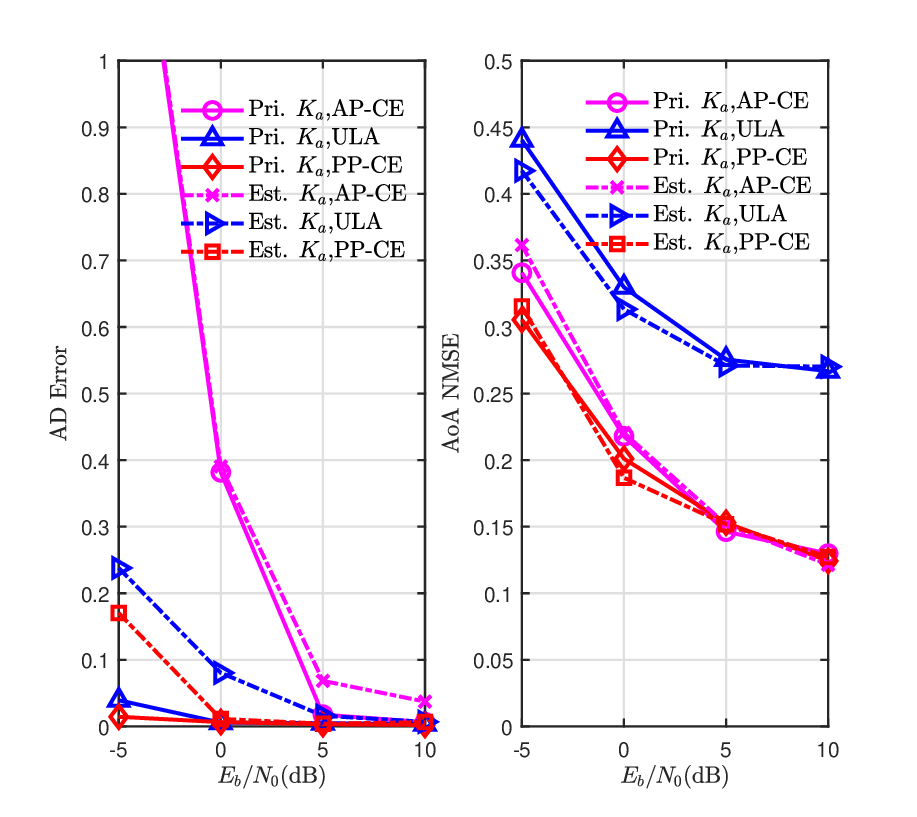}
	\caption{Performance of AD And AoA NMSE with/without activity prior.}
	\label{ADE_AOA_NMSE}
\end{figure}
We illustrate the performance of AD and AoA estimation with $K_a$ prior and with estimated $\tilde{K}_a$ by \eqref{eq:6} under different estimation strategy and channel model in Fig. \ref{ADE_AOA_NMSE} with different energy-per-bit. Overall, the precision of estimation drops with the increased energy-per-bit. Moreover, the proposed PP-CE shows the best performance among all in terms of AD error and AoA NMSE. It indicates that FAS-URA has potential to obtain enhanced estimation capability compared with conventional ULA channel model. Notably, though the proposed AP-CE only utilizes shortened pilot, it has much better AoA estimation performance compared with ULA model. It means the proposed AP-CE has more superiority in scenario where AoA estimation matters.
\begin{figure}[htp]
	\centering
	\includegraphics[width=3.5in]{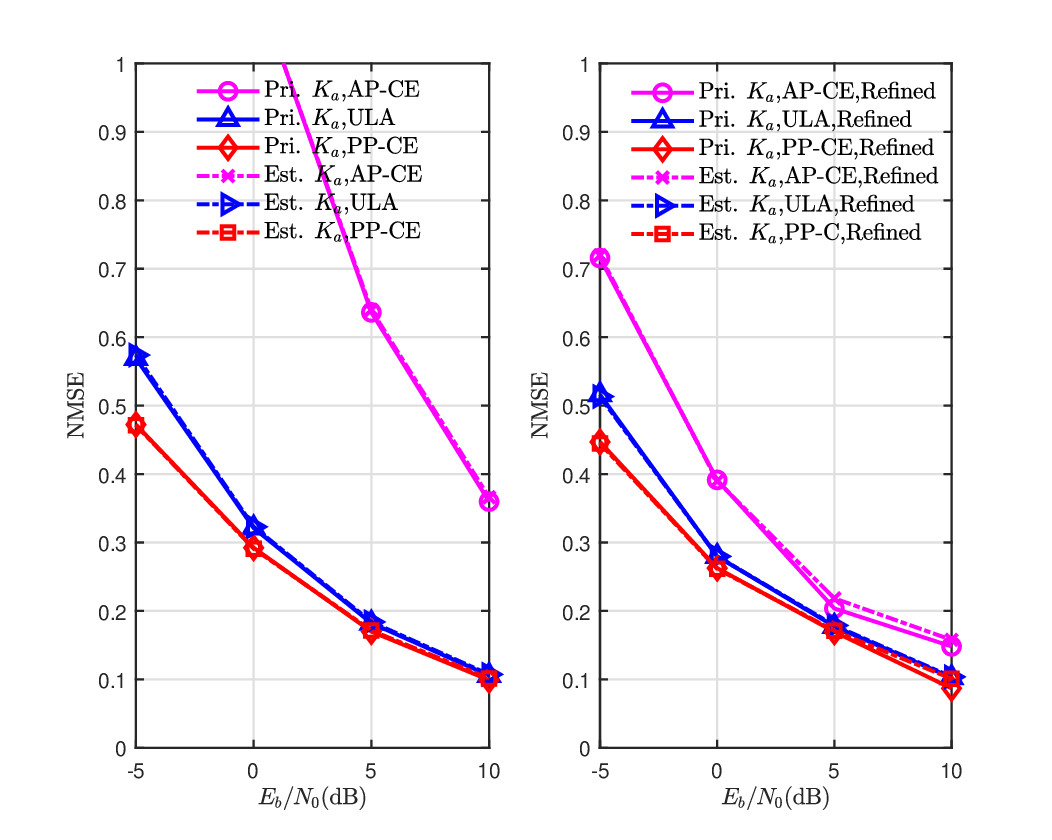}
	\caption{Performance of CE with refinement with/without activity prior.}
	\label{NMSE_before_refine_bnr}
\end{figure}
\subsubsection{CE and Estimation Refinement}
Furthermore, we compare the performance of CE with refinement via geometric models of FAS and ULA. Overall, the proposed PP-CE has the best CE performance compared with others with or without estimation refinement by channel dictionary. Notably, CE refinement can drastically improve the performance of the proposed AP-CE. Meanwhile, the prior of activity does not generate much derivation for CE under all strategies or models. 
\begin{figure}[htp]
	\centering
	\includegraphics[width=3.8in]{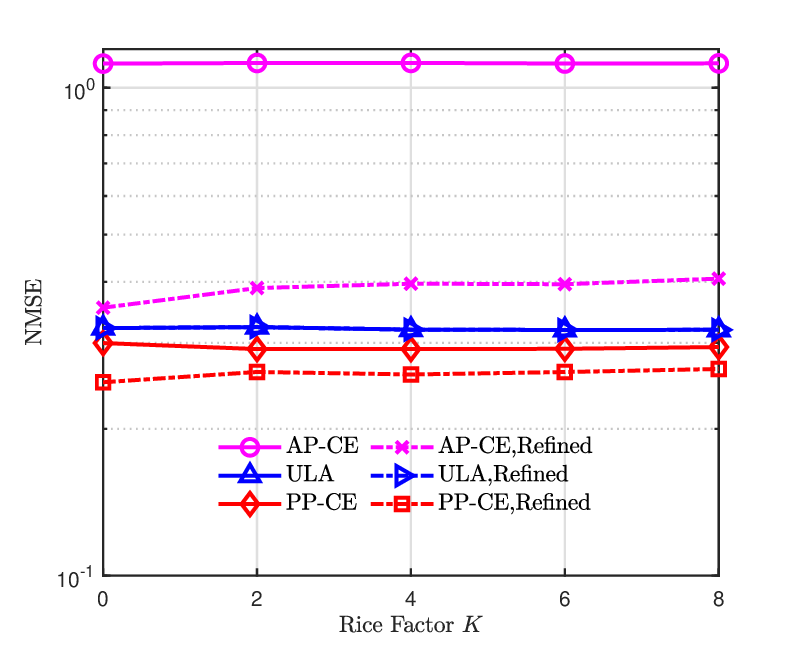}
	\caption{Performance of CE versus different Rice factor with $E_b/N_0=0$dB.}
	\label{NMSE_Rice_Factor}
\end{figure}
We also investigate the impact from LOS path with different Rice factor with $E_b/N_0=0$dB. The results demonstrate that the proposed schemes can well tackle CE under different Rice factor.
\section{Conclusion}\label{sec.7}
In this work, we investigate the CE problem for URA with fluid antennas. Channel coefficient and AoA estimation are enabled through different strategies, namely alternate/partial ports CE, by exploiting the sparsity of finite scatterers. In particular, a regularized estimator for partial ports CE is developed, with an optimized port index gap, to achieve desirable AoA estimation and improved channel refinement. Numerical results highlight the promising potential of the FAS-URA. Future work will address concatenated decoding, considering FAS diversity.
\section*{Acknowledgment}
This work is supported by NSFC projects (61971136,
61960206005), the Fundamental Research Funds for the
Central Universities (2242022k60001, 2242021R41149,
2242023K5003).
\balance

%\vspace{12pt}
%\color{red}
%IEEE conference templates contain guidance text for composing and formatting conference papers. Please ensure that all template text is removed from your conference paper prior to submission to the conference. Failure to remove the template text from your paper may result in your paper not being published.

\end{document}